\documentclass[10pt]{article}
\usepackage{amsmath}
\usepackage{amssymb}
\usepackage{graphicx}
\usepackage{cite}
\usepackage{geometry}
\usepackage{color}
\usepackage[labelfont=bf,labelsep=period,justification=raggedright]{caption}
\makeatletter
\renewcommand{\@biblabel}[1]{\quad#1.}
\makeatother

\date{}

\begin{document}

\begin{flushleft}
{\Large
\textbf{Phenotypic Heterogeneity in Mycobacterial Stringent Response}
}
\\
Sayantari Ghosh$^{1}$, 
Kamakshi Sureka$^{2}$, 
Bhaswar Ghosh$^{3}$,
Indrani Bose$^{1\ast}$,
Joyoti Basu$^{2}$,
Manikuntala Kundu$^{2}$

\scriptsize{\textbf{1} Department of Physics, Bose Institute, Kolkata, India,
\textbf{2} Department Of Chemistry, Bose Institute, Kolkata, India,
\textbf{3} Centre for Applied Mathematics and Computational Science , 
           Saha Institute of Nuclear Physics, Kolkata, India}
\\
\scriptsize{$\ast$ E-mail: indrani@bosemain.boseinst.ac.in}
\end{flushleft}

\begin{abstract}
        % Do not use inserted blank lines (ie \\) until main body of text.
        \textbf{Background:} A common survival strategy of microorganisms subjected
to stress involves the generation of phenotypic heterogeneity in the
isogenic microbial population enabling a subset of the population
to survive under stress. In a recent study, a mycobacterial population
of \textit{M. smegmatis} was shown to develop phenotypic heterogeneity
under nutrient depletion. The observed heterogeneity is in the form
of a bimodal distribution of the expression levels of the Green Fluorescent
Protein (GFP) as reporter with the \textit{gfp} fused to the promoter
of the\textit{ rel }gene. The stringent response pathway is initiated
in the subpopulation with high \textit{rel} activity.
      
        \textbf{Results:} In the present study, we characterise quantitatively the
single cell promoter activity of the three key genes, namely, \textit{mprA,
sigE} and \textit{rel}, in the stringent response pathway with \textit{gfp
}as the reporter. The origin of bimodality in the GFP distribution
lies in two stable expression states, i.e., bistability. We develop
a theoretical model to study the dynamics of the stringent response
pathway. The model incorporates a recently proposed mechanism of bistability
based on positive feedback and cell growth retardation due to protein
synthesis. Based on flow cytometry data, we establish that the distribution
of GFP levels in the mycobacterial population at any point of time
is a linear superposition of two invariant distributions, one Gaussian
and the other lognormal, with only the coefficients in the linear
combination depending on time. This allows us to use a binning algorithm
and determine the time variation of the mean protein level, the fraction
of cells in a subpopulation and also the coefficient of variation,
a measure of gene expression noise.

        \textbf{Conclusions:}  The results of the theoretical model along with a comprehensive
analysis of the flow cytometry data provide definitive evidence for
the coexistence of two subpopulations with overlapping protein distributions.

\end{abstract}

\section*{Background}

Microorganisms are subjected to a number of stresses during their
lifetime. Examples of such stresses are: depletion of nutrients, environmental
fluctuations, lack of oxygen, application of antibiotic drugs etc.
Microorganisms take recourse to a number of strategies for survival
under stress and adapting to changed circumstances \cite{1,2,3,4}. A prominent
feature of such strategies is the generation of phenotypic heterogeneity
in an isogenic microbial population. The heterogeneity is advantageous
as it gives rise to variant subpopulations which are better suited
to persist under stress. Bistability refers to the appearance of two
subpopulations with distinct phenotypic characteristics \cite{5,6}.
In one of the subpopulations, the expression of appropriate stress
response genes is initiated resulting in adaptation . There are broadly
two mechanisms for the generation of phenotypic heterogeneity \cite{7,8}.
In ``responsive switching'' cells switch
phenotypes in response to perturbations associated with stress. In
the case of ``spontaneous stochastic switching'',
transitions occur randomly between the phenotypes even in the absence
of stress. Responsive switching may also have a stochastic component
as fluctuations in the level of a key regulatory molecule can activate
the switch once a threshold level is crossed \cite{1,3,5,6}.

The pre-existing phenotypic heterogeneity, an example of the well-known
``bet-hedging-strategy\textquotedblright, keeps the population
in readiness to deal with future calamities. Using a microfluidic
device, Balaban et al. \cite{9} have demonstrated the existence of
two distinct subpopulations, normal and persister, in a growing colony
of \textit{E. coli} cells. The persister subpopulation constitutes
a small fraction of the total cell population and is distinguished
from the normal subpopulation by a reduced growth rate. Since killing
by antibiotic drugs like ampicillin depends on the active growth of
cell walls, the persister cells manage to survive when the total population
is subjected to antibiotic treatment. The normal cells, with an enhanced
growth rate are, however, unable to escape death. Once the antibiotic
treatment is over, some surviving cells switch from the persister
to the normal state so that normal population growth is resumed \cite{9,11}.
A simple theoretical model involving transitions between the normal
and persister phenotypes explains the major experimental observations
well \cite{9,10}. In the case of environmental perturbations, Thattai
and Oudenaarden \cite{12} have shown through mathematical modelling
that a dynamically heterogeneous bacterial population can under certain
circumstances achieve a higher net growth rate conferring a fitness
advantage than a homogeneous one. Mathematical modelling further shows
that responsive switching is favoured over spontaneous switching in
the case of rapid environmental fluctuations whereas the reverse is
true when environmental perturbations are infrequent \cite{7}. Another
theoretical prediction that cells may tune the switching rates between
phenotypes to the frequency of environmental changes has been verified
in an experiment by Acar et al. \cite{13} involving an engineered strain
of S. cerevisiae which can switch randomly between two phenotypes.
The major feature of all such studies is the coexistence of two distinct
subpopulations in an isogenic population and their interconversions
in the presence/absence of stress. Bistability, i.e., the partitioning
of a cell population into two distinct subpopulations has been experimentally
observed in a number of cases \cite{1,3,5}. Some prominent examples
include: lysis/lysogeny in bacteriophage $\lambda$ \cite{14}, the
activation of the lactose utilization pathway in \textit{E. coli } \cite{15}
and the galactose utilization genetic circuit in S. cerevisiae \cite{16},
competence development in \textit{B. subtilis} \cite{6,17,18} and the
stringent response in mycobacteria \cite{19}.

The mycobacterial pathogen \textit{M. tuberculosis}, the causative
agent of tuberculosis, has remarkable resilience against various physiological
and environmental stresses including that induced by drugs \cite{20,21,22}.
On tubercular infection, granulomas form in the host tissues enclosing
the infected cells. Mycobacteria encounter a changed physical environment
in the confined space of granulomas with a paucity of life-sustaining
constituents like nutrients, oxygen and iron \cite{23,24}. The pathogens
adapt to the stressed conditions and can survive over years in the
so-called latent state. In vitro too, \textit{M. tuberculosis} has
been found to persist for years in the latent state characterised
by the absence of active replication and metabolism \cite{25}. Researchers
have developed models simulating the possible environmental conditions
in the granulomas. One such model is the adaptation to nutrient-depleted
stationary phase \cite{26}. The processes leading to the slowdown of
replicative and metabolic activity constitute the stringent response.
In mycobacteria, the expression of \textit{rel }initiates the stringent
response which leads to persistence. The importance of Rel arises
from the fact that it synthesizes the stringent response regulator
ppGpp (guanosine tetraphosphate) \cite{27} and is essential for the
long-term survival of \textit{M. tuberculosis} under starvation \cite{28}
and for prolonged life of the bacilli in mice \cite{29}.

Key elements of the stringent response and the ability to survive
over long periods of time under stress are shared between the mycobacterial
species \textit{M. tuberculosis} and \textit{M. smegmatis} \cite{30}.
Recent experiments provide knowledge of the stress signaling pathway
in mycobacteria linking polyphosphate (poly P), the two-component
system MprAB, the alternate sigma factor SigE and Rel \cite{31}. In
an earlier study \cite{19}, we investigated the dynamics of \textit{rel-gfp
}expression (\textit{gfp} fused with \textit{rel} promoter) in \textit{M.
smegmatis} grown upto the stationary phase with nutrient depletion
serving as the source of stress. In a flow cytometry experiment, we
obtained evidence of a bimodal distribution in GFP levels and suggested
that positive feedback in the stringent response pathway and gene
expression noise are responsible for the creation of phenotypic heterogeneity
in the mycobacterial population in terms of the expression of \textit{rel-gfp}.
Positive feedback gives rise to bistability \cite{5,6}, i.e., two stable
expression states corresponding to low and high GFP levels. We further
demonstrated hysteresis, a feature of bistability, in \textit{rel-gfp}
expression. The mathematical model developed by us to study the dynamics
of the stringent response pathway predicted bistability in a narrow
parameter regime which, however, lacks experimental support. In general,
to obtain bistability a gene circuit must have positive feedback and
cooperativity in the regulation of gene expression. Recently, Tan
et al. \cite{32} have proposed a new mechanism by which bistability
arises from a noncooperative positive feedback circuit and circuit-induced
growth retardation. The novel type of bistability was demonstrated
in a synthetic gene circuit. The circuit, embedded in a host cell,
consists of a single positive feedback loop in which the protein product
X of a gene promotes its own synthesis in a noncooperative fashion.
The protein decay rate has two components, the degradation rate and
the dilution rate due to cell growth. In the circuit considered, production
of X slows down cell growth so that at higher concentrations of X,
the rate of dilution of X is reduced. This generates a second positive
feedback loop since increased synthesis of X proteins results in faster
accumulation of the proteins so that the protein concentration is
higher. The combination of two positive feedback loops gives rise
to bistability in the absence of cooperativity. A related study by
Klumpp et al \cite{35} has also suggested that cell growth inhibition
by a protein results in positive feedback.

%%%%%%%%%%%%%%%%%%%%%%%%%%%%
%% Results and Discussion %%
\section*{Results}
In this paper, we develop a theoretical model incorporating the effect
of growth retardation due to protein synthesis \cite{32,33}. We provide
some preliminary experimental evidence in support of the possibility.
In our earlier study \cite{19}, bimodality in the \textit{rel-gfp}
expression levels was observed. As a control, GFP
expression driven by the constitutive \textit{hsp60}
promoter was monitored as a function of time. A single bright population
was observed at different times of growth (Figure S4 of \cite{19}).
The unimodal rather than bimodal distribution ruled out the possibility
that clumping of mycobacterial cells and cell-to-cell variation of
plasmid copy number were responsible for the observed bimodal fluorescence
intensity distribution of \textit{rel}
promoter driven GFP expression. In the present study, we perform
flow cytometry experiments to monitor \textit{mprA-gfp} and \textit{sigE-gfp}
expression levels. The distribution of GFP levels in each case is
found to be bimodal. We determine the probability distributions of
the two subpopulations associated with low and high expression levels
at different time points in the three cases of \textit{mprA-gfp},
\textit{sigE-gfp} and \textit{rel-gfp} expression. In each case, the total
distribution is a linear combination of two invariant distributions
with the coefficients in the linear combination depending on time.
The results of hysteresis experiments are also reported.

\subsection*{Mathematical modeling of the stress response pathway}

\begin{figure}
\begin{center}
\includegraphics[scale=0.5]{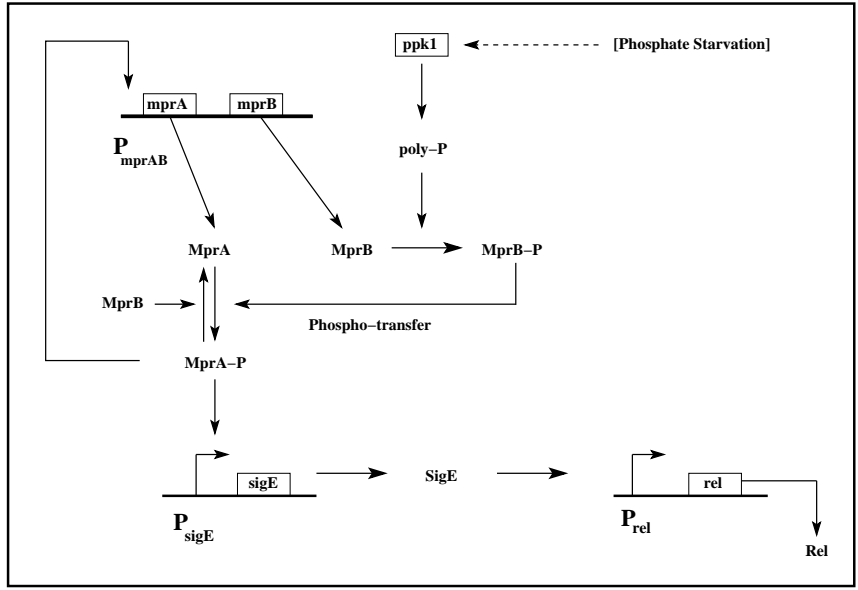}
\end{center}
\caption{\textbf{Schematic diagram of the stringent response
pathway in }\textbf{\textit{M. smegmatis}}\textbf{ activated under
nutrient depletion.} MprB-P and MprA-P are the phosphorylated forms
of MprB and MprA respectively. Poly P serves as the phosphate donor
in the conversion of MprB to MprB-P.} 
\end{figure}

Figure 1 shows a sketch of the important components of the stress
response pathway in \textit{M. smegmatis} subjected to nutrient depletion
\cite{19,31}. The operon \textit{mprAB} consists of two genes \textit{mprA}
and \textit{mprB} which encode the histidine kinase sensor MprB and
its partner the cytoplasmic response regulator MprA respectively.
The protein pair responds to environmental stimuli by initiating adaptive
transcriptional programs. Polyphosphate kinase 1 (PPK1) catalyses
the synthesis of polyphosphate (poly P) which is a linear polymer
composed of several orthophosphate residues. Mycobacteria
possibly encounter a phosphate-limited environment in macrophages.
Sureka et al. \cite{31} proposed that poly P could play a critical
role under ATP depletion by providing phosphate for utilisation by
MprAB. A recent experiment \cite{33} on a population of \emph{M.
tuberculosis} has established that the MTB gene
\emph{ppk1} is significantly upregulated
due to phosphate starvation resulting in the synthesis of inorganic
poliphosphate (poly P). The two-component regulatory system SenX3-RegX3
is known to be activated on phosphate starvation in both \emph{M.
smegmatis } \cite{34} and \emph{M.
tuberculosis} \cite{33}. In the latter case, RegX3
has been shown to regulate the expression of \emph{ppk1},
a feature expected to be shared by \emph{M. smegmatis}.
In both the mycobacterial populations, poly P regulates the stringent
response via the \emph{mprA-sigE-rel}
pathway \cite{31}. In our experiments, nutrient depletion possibly
gives rise to phosphate starvation. On activation
of the \emph{mprAB} operon,
MprB autophosphorylates itself with poly P serving as the phosphate
donor \cite{31,33}. The phosphorylated MprB-P phosphorylates MprA via
phosphotransfer reactions. There is also evidence that MprB functions
as a MprA-P (phosphorylated MprA) phosphatase. MprA-P binds the promoter
of the \textit{mprAB} operon to initiate transcription. A positive
feedback loop is functional in the signaling network as the production
of MprA brings about further MprA synthesis. The \textit{mprAB} operon
has a basal level of gene expression independent of the operation
of the positive feedback loop. Once the \textit{mprAB} operon is activated,
MprA-P regulates the transcription of the alternate sigma factor gene
\textit{sigE}, which in turn controls the transcription of \textit{rel}.
We construct a mathematical model to study the dynamics of the above
signaling pathway. The new feature included in the model takes into
account the possibility that the production of stress-induced proteins
like MprA and MprB slows down cell growth. This effectively generates
a positive feedback loop as explained in Refs. \cite{32,33}.
\begin{figure}
\begin{center}
\includegraphics[scale=0.6]{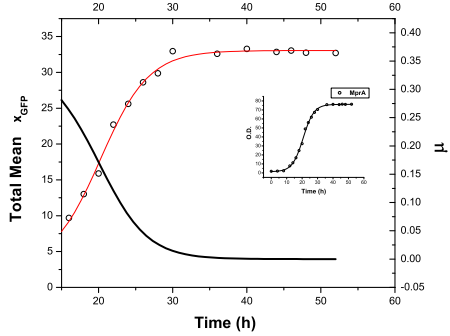}
\end{center}
\caption{\textbf{Growth retardation due to protein synthesis.}
(a) Mean amount of GFP fluorescence in the case of \textit{mprA} promoter
fused with \textit{gfp} and (b) specific growth rate $\mu$ of mycobacterial
population versus time in hours (h).} 
\end{figure}
\textbf{
}Figure 2(a) shows the mean amount of GFP fluorescence in the total
mycobacterial population as measured in a flow cytometry experiment
(\textit{mprA }promoter fused with\textit{ gfp}) versus time. Figure
2(a) shows the specific growth rate of the cell population versus
time. The inset shows the experimental growth curve for the mycobacterial
population. The growth was monitored by recording the absorbance values
at 600 nm spectrophotometrically (see Methods). The specific growth
rate at time $t$ is given by $\frac{1}{N(t)}\frac{dN(t)}{dt}$
where $N(t)$ is the number of mycobacterial cells at time $t$. Nutrient
depletion limits growth and proliferation and culminates in the activation
of stress response genes. It appears that in many cases rapid growth
and stress response are mutually exclusive so that the production
of a stress response protein gives rise to a slower growth rate \cite{36}.
The balance between the expressions of growth-related and stress-induced
genes determines the cellular phenotype with respect to growth rate
and stress response. Persister cells in both \textit{E. coli} \cite{9,11}
and mycobacteria \cite{21,22} have slow growth rates. In the case of
\textit{M. smegmatis}, we have already established that the slower
growing persister subpopulation has a higher level of Rel, the initiator
of stringent response, as compared to the normal subpopulation \cite{19}.
The new addition to our mathematical model \cite{19} involves nonlinear
protein decay rates arising from cell growth retardation due to protein
synthesis. We briefly discuss the possible origins of the nonlinearity
and its mathematical form \cite{32,33}. The temporal rate of change
of protein concentration is a balance between two terms: rate of synthesis
and rate of decay. The decay rate constant $(\gamma_{eff})$ has two
components: the dilution rate due to cell growth ($\mu$) and the
natural decay rate constant ($\gamma$), i.e., $\gamma_{eff}=\mu+\gamma$
where $\mu$ is the specific growth rate. In many cases, the expression
of a protein results in cell growth retardation \cite{32,33}. The
general form of the specific growth rate in such cases is given by
\begin{equation}
\mu=\frac{\phi}{1+\theta x}\end{equation}

where $x$ denotes the protein concentration and $\phi,\theta$ are
appropriate parameters. In Ref. \cite{32}, the expression for $\mu$
(Eq. (1)) is arrived at in the following manner. The Monod model \cite{37}
takes into account the effect of resource or nutrient limitation on
the growth of bacterial cell population. The rate of change in the
number of bacterial cells is

\begin{equation}
\frac{dN}{dt}=\mu N\end{equation}
where the specific growth rate $\mu$ is given by

\begin{equation}
\mu=\mu_{max\:}\frac{s}{k+s}\end{equation}
In (3), $s$ is the nutrient concentration and $k$ the half saturation
constant for the specific nutrient. When $s=k$, the specific growth
rate attains its half maximal value ($\mu_{max}$ is the maximum value
of specific growth rate). The metabolic burden of protein synthesis
affecting the growth rate is modeled by reducing the nutrient amount
$s$ by $\epsilon$, i.e.,

\begin{equation}
\mu=\frac{\mu_{max}}{1+\frac{k}{s(1-\epsilon)}}\end{equation}

The magnitude of $\epsilon$ is assumed to be small and proportional
to the protein concentration $x$. Following the procedure outlined
in the Supplementary Information of \cite{32}, namely, applying Taylor's
expansion to (4) and putting $\epsilon=\lambda x$ ($\lambda$ is
a constant), one obtains the expression in Eq. (1) with $\phi\mbox{ = }\frac{\mu_{max}s}{s+k}$
and $\theta=\frac{k\lambda}{s+k}$ . Thus, the decay rate of proteins
has the form $-\gamma_{eff}x=-(\gamma+\mu)\, x$ where $\mu$ is given
by Eq. (1). There are alternative explanations for the origin of the
nonlinear decay term, e.g., the synthesis of a protein may retard
cell growth if it is toxic to the cell \cite{35}. In the case of mycobacteria,
there is some experimental evidence of cell growth retardation brought
about by protein synthesis. The response regulator MprA has an essential
role in the stringent response pathway leading to persistence of mycobacteria
under nutrient deprivation. Inactivation of the regulator in an \textit{mprA}
insertion mutant resulted in reduced persistence in a murine model
but the growth of the mutant was proved to be significantly higher
than that observed in the cases of the wild-type species \cite{38,39}.

\begin{figure}
\begin{center}
\includegraphics[scale=1]{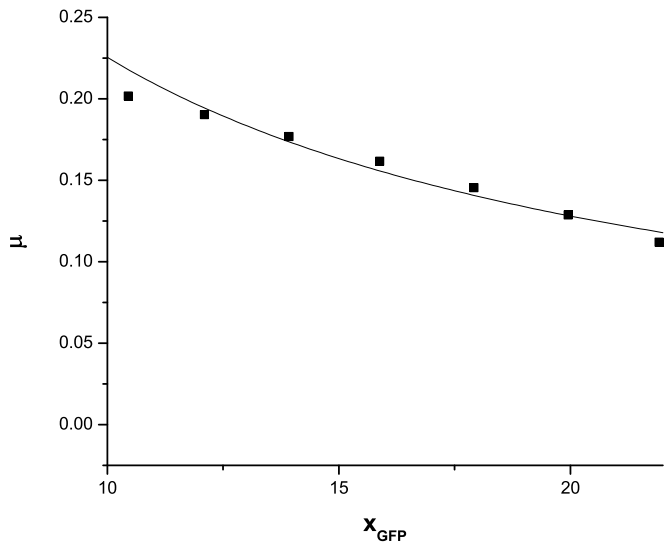}
\end{center}
\caption{\textbf{Specific growth rate $\mu$ versus GFP fluorescence
intensity $x_{GFP}$ fitted with an expression similar to that given
in Eq. (1). } The values of $\mu_{max}^{GFP}$ and $\theta^{GFP}$ are $\mu_{max}^{GFP}=0.94$
and $\theta^{GFP}=0.317$. The data points correspond to the growth
period of 16-23 hours.} 
\end{figure}

Our experimental data (Figure 3) provide further
support to the hypothesis that MprA synthesis leads to reduced specific
growth rate. The data points represent GFP fluorescence intensity
with \textit{gfp } fused to the
\emph{mprA} promoter. The GFP
acts as a reporter of the \textit{mprA}
promoter activity culminating in MprA (also MprB) synthesis.
The data points shown in Figure 3 are those that correspond to the
growth period of 16-23 hours in Figure 2.

The data points are fitted by an expression similar to that in Eq.
(1) with $\mu_{max}^{GFP}=0.94$ and $\theta^{GFP}=0.317$. The differential
equations describing the temporal rates of change of key protein concentrations
in our model are described in the Additional File 1. Solving the equations,
one finds the existence of bistability, i.e., two stable expression
states in an extended parameter regime. Figures S1 A-C (Additional
File 1) show the plots for bistability and hysteresis for the proteins
MprA, SigE and Rel versus the autophosphorylation rate. In the deterministic
scenario and in the bistable regime, all the cells in a population
are in the same steady state if exposed to the same environment and
with the same initial state. The experimentally observed heterogeneity
in a genetically identical cell population is a consequence of stochastic
gene expression. The biochemical events involved in gene expression
are inherently probabilistic \cite{40,41} in nature. The uncertainty
introduces fluctuations (noise) around mean expression levels so that
the single protein level of the deterministic case broadens into a
distribution of levels. In the case of bistable gene expression, the
distribution of protein levels in a population of cells is bimodal
with two distinct peaks.

\subsection*{Bimodal Expression of \textit{mprA}, \textit{sigE} and \textit{rel}
in \textit{M. smegmatis}}

In the earlier study \cite{19}, we investigated the dynamics of \textit{rel
}transcription in individual cells of \textit{M. smegmatis} grown
in nutrient medium up to the stationary phase, with nutrient depletion
serving as the source of stress. We employed flow cytometry to monitor
the dynamics of green fluorescent protein (GFP) expression in \textit{M.
smegmatis} harboring the \textit{rel} promoter fused to \textit{gfp
}as a function of time. The experimental signature of bistaility lies
in the coexistence of two subpopulations. We now extend the study
to investigate the dynamics of \textit{mprA} and \textit{sigE} transcription
in individual \textit{M. smegmatis} cells in separate flow cytometry
experiments. Figures 4(a) and (b) show the time course of \textit{mprA}-GFP
and \textit{sigE}-GFP expressions respectively as monitored by flow
cytometry. In both the cases, the distribution of GFP-expressing cells
is bimodal indicating the existence of two distinct subpopulations.
In each case, the cells initially belong to the subpopulation with
low GFP expression. The fraction of cells with high GFP expression
increases as a function of time. The two subpopulations with low and
high GFP expression are designated as L and H subpopulations respectively.
In the stationary phase, the majority of the cells belong to the H
subpopulation. The presence of two distinct subpopulations confirms
the theoretical prediction of bistability.

\begin{figure}
\begin{center}
\includegraphics[scale=1]{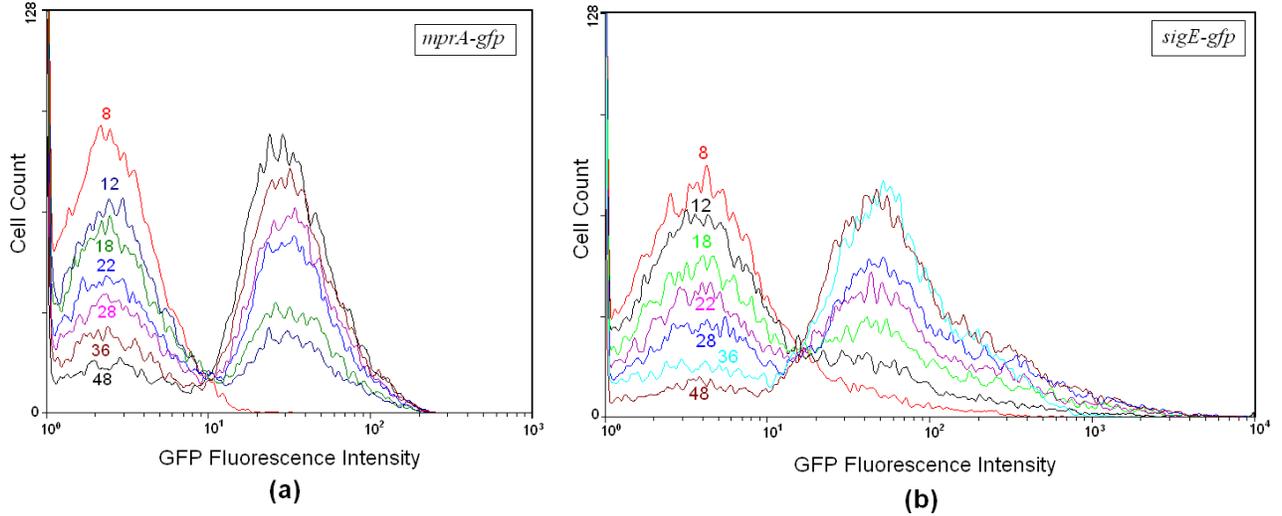}
\end{center}
\caption{\textbf{Time course of (a) }\textbf{\textit{mprA-gfp}}\textbf{
and (b) }\textbf{\textit{sigE-gfp}}\textbf{ expression.} \textit{M.
smegmatis} harboring the appropriate promoter construct was grown
for different periods of time (indicated in hours (h)) and the specific
promoter-driven expression of GFP was monitored by flow cytometry.
With time, there is a gradual transition from the L to the H subpopulation.} 
\end{figure}

We analysed the experimental data shown in Figure 4 and found that
at any time point the distribution $P(x,t)$ of GFP levels in a population
of cells is a sum of two overlapping and time-independent distributions,
one Gaussian ($P_{1}(x)$) and the other lognormal ($P_{2}(x)$),
i.e.,

\begin{center}
\begin{equation}
P(x,t)=C_{1}(t)P_{1}(x)+C_{2}(t)P_{2}(x)\end{equation}

\par\end{center}

The coefficients $C_{i}$'s (i=1, 2) depend on time whereas $P_{1}(x)$
and $P_{2}(x)$ are time-independent. The general forms of $P_{1}(x)$
and $P_{2}(x)$ are,

\begin{equation}
P_{1}(x)=\frac{exp(-(\frac{x-x_{01}}{w_{01}})^{2})}{w_{01}\,\sqrt{\frac{\pi}{2}}}\end{equation}

\begin{equation}
P_{2}(x)=\frac{exp(-\frac{1}{2}(\frac{lnx-x_{02}}{w_{02}})^{2})}{x\, w_{02}\,\sqrt{2\pi}}\end{equation}

Figures S2(a) and (b) in Additional File 1 illustrate the typical
forms of the Gaussian and lognormal distributions. The Gaussian distribution
has a symmetric form whereas the lognormal distribution is asymmetric
and long-tailed.
\begin{figure}
\begin{center}
\includegraphics[scale=1]{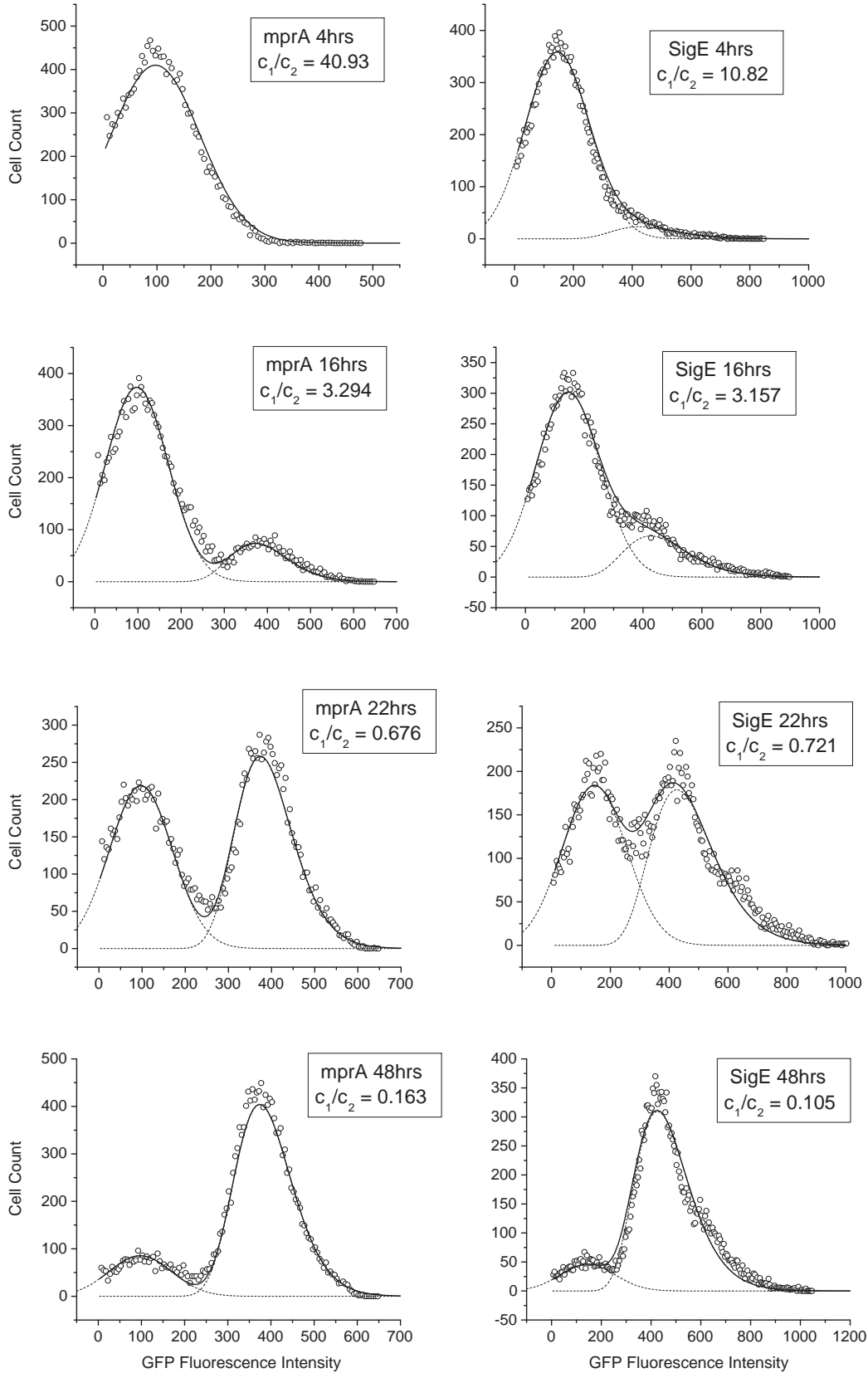}
\end{center}
\caption{\textbf{Fitting of data with two distributions.}
Experimental data for cell count versus GFP fluorescence intensity
at selected time points when \textit{gfp }is fused with \textit{mprA}
and \textit{sigE }promoters respectively. The solid curve represents
$P(x,t)$ in equation (5) and the dotted curves are the individual
terms on the r.h.s.} 
\end{figure}
Figure 5 shows the experimental data for cell count
versus GFP fluorescence intensity at selected time points in the cases
when\textit{ gfp} is fused with \textit{mprA} and \textit{sigE} promoters
in separate experiments. The dotted curves represent the individual
terms in the r.h.s. of Eq. (5) and the solid curve denotes the linear
combination $P(x,t)$. The different parameters of $P_{1}(x)$ and
$P_{2}(x)$ have the values $x_{01}=97.3366\:(145.86181),w_{01}=103.0731\:(154.67381),x_{02}=5.95526\:(6.1171),w_{02}=0.17618\:(0.2509)$
when \textit{gfp} is fused with \textit{mprA} (\textit{sigE}). The
ratio of the coefficients, $C_{1}(t)/C_{2}(t)$, has the value listed
by the side of each figure. Figure S3 displays a similar analysis
of the experimental data when \textit{gfp} is fused to the \textit{rel}
promoter.

In the earlier study \cite{19}, the total cell population was divided
into L and H subpopulations depending on whether the measured GFP
fluorescence intensity was less or greater than a threshold intensity.
In the present study, we have obtained approximate analytic expressions
for the distributions of GFP fluorescence intensity in the L and H
subpopulations. The two distributions, Gaussian and lognormal, have
overlaps in a range of fluorescence intensity values (Figure 5 and
Figure S3 in Additional File 1). We next used the binning algorithm
developed by Chang et al. \cite{42} to partition the cells of the total
population into two overlapping distributions, one Gaussian (Eq. (6))
and the other lognormal (Eq. (7)). At time t, let N(t) be the total
number of cells. For each cell, the data $x_{j}$ for the fluorescence
intensity is used to calculate the ratios,

\begin{equation}
g_{1}(x_{j})=\frac{P_{1}(x_{j})}{P_{1}(x_{j})+P_{2}(x_{j})}\:,\: g_{2}(x_{j})=\frac{P_{2}(x_{j})}{P_{1}(x_{j})+P_{2}(x_{j})}\end{equation}

where $P_{1}(x)$ and $P_{2}(x)$ are the distributions in Eqs. (6)
and (7). A random number r is generated and the cell $j$ is assigned
to the L subpopulation if $0\leq r<g_{1}(x_{j})$ , the cell belongs
to the H subpopulation otherwise. Once the total population is partitioned
into the L and H subpopulations, one can calculate the following quantities:

\begin{eqnarray}
\omega_{i}(t) & = & \frac{N_{i}(t)}{N(t)}\;\;\;(i=1,2) \nonumber \\
\mu_{i}(t) & = & \sum x_{ji}(t)/N_{i}(t) \\
\sigma_{i}^{2}(t) & = &\sum(x_{ji}(t)-\mu_{i}(t))^{2}/(N_{i}(t)-1) \nonumber 
\end{eqnarray}

The indices $i=1,2$ correspond to the L and H subpopulations, $\omega_{i}(t)$
is the fraction of cells in the $i$th subpopulation at time t, $\mu_{i}(t)$
is the mean fluorescence intensity for the $i$th subpopulation and
$\sigma_{i}^{2}(t)$ the associated variance.
\begin{figure}
\includegraphics[scale=1]{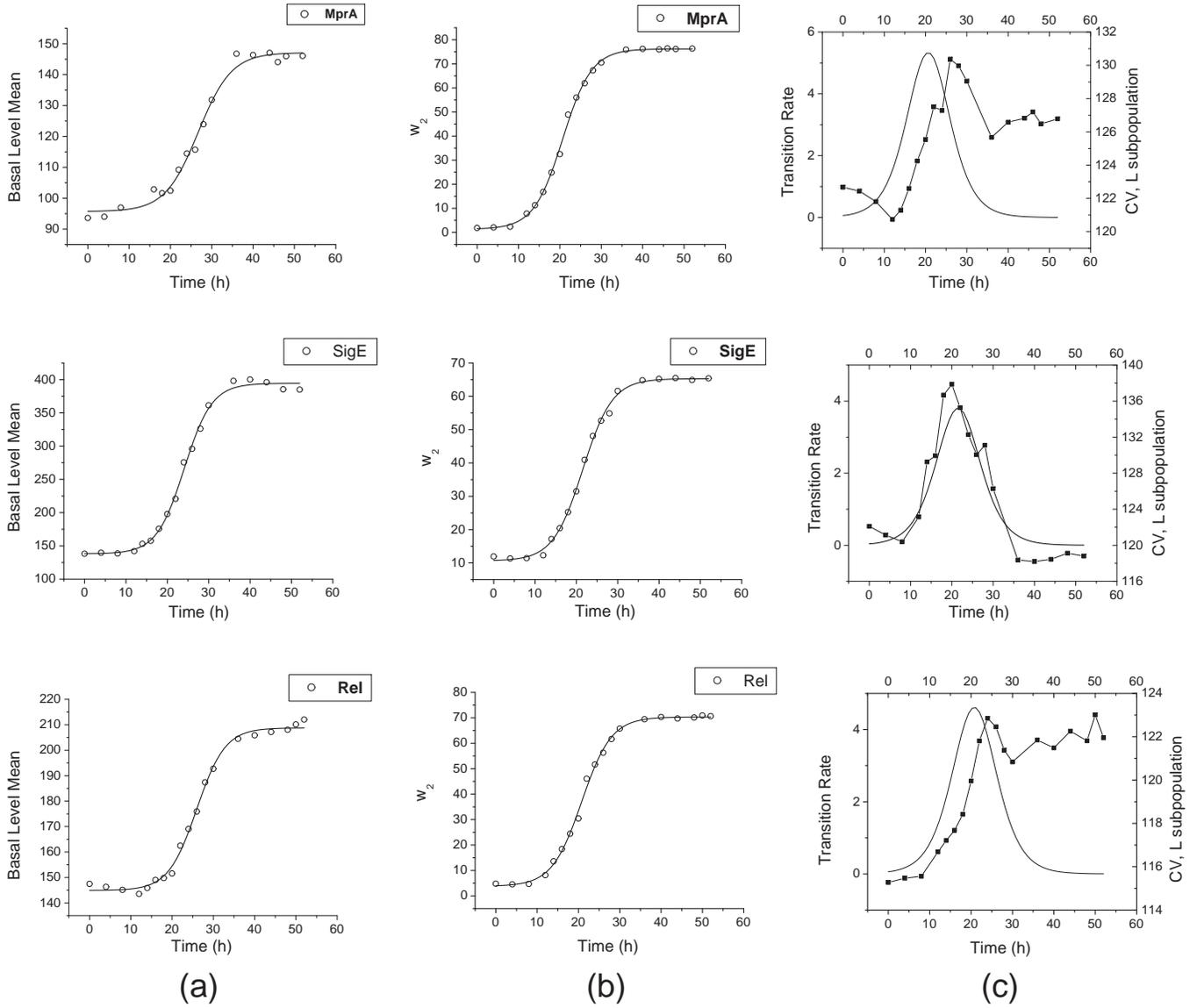}
\caption{\textbf{ Analysis of the time course of gfp expression.
}(a) Mean protein level in L subpopulation (basal level) versus time
in hours in the three cases of \textit{gfp} fused with \textit{mprA,
sigE }and\textit{ rel} promoters respectively. (b) Fraction of cells
$\omega_{2}$(t) in the H subpopulation versus time in hours in the
three cases. (c) Transition rate from the L to the H subpopulation
and the CV of the protein levels in the L subpopulation versus time
in hours in the three cases. The experimental data are analysed using
binning algorithm to obtain the plots (a), (b) and (c).} 
\end{figure}
Figure 6 shows the results
of the data analysis. Figure 6(a) shows the plots of mean GFP fluorescence
level for the L subpopulation (basal level) versus time in the three
cases of\textit{ gfp }fused with the promoters of\textit{ mprA}, \textit{sigE}
and \textit{rel} respectively. Figure 6(b) displays the data for the
fractions of cells, $\omega_{2}(t)$ , versus time in the three cases
and Figure 6(c) shows the transition rate versus time along with the
coefficients of variation CV (CV= standard deviation/mean) of the
protein levels in the L subpopulation versus time.

Figure S4 (Additional File 1) shows the plots of mean GFP fluorescence
level for the total population versus time in the three cases of \textit{gfp}
fused with the promoters of \textit{mprA}, \textit{sigE} and \textit{rel}
respectively. As in the case of the basal level versus time data (Figure
6(a)), the plots are sigmoidal in nature. We solved the differential
equations of the theoretical model described in Additional File 1
and obtained the concentrations of MprA, MprB, SigE, MprA-P, MprB-P
and GFP versus time. Some of these plots are shown in Figure S5 (Additional
File 1) and reproduce the sigmoidal nature of the experimental plots.
We note that the sigmoidal nature of the curves is obtained only when
the non-linear nature of the degradation rate is taken into account.

As we have already discussed, the distribution of
GFP levels in the mycobacterial cell population is a linear combination
of two invariant distributions, one Gaussian and the other lognormal,
with only the coefficients in the linear combination dependent on
time. Friedman et al. \cite{43} have developed an analytical framework
of stochastic gene expression and shown that the steady state distribution
of protein levels is given by the gamma distribution. The theory was
later extended to include the cases of transcriptional autoregulation
as well as noise propagation in a simple genetic network. While experimental
support for gamma distribution has been obtained earlier \cite{44},
a recent exhaustive study \cite{45} of the \emph{
E. coli} proteome and transcriptome with single-molecule
sensitivity in single cells has established that the distributions
of almost all the protein levels out of the 1018 proteins investigated,
are well fitted by the gamma distribution in the steady state. The
gamma distribution was found to give a better fit than the lognormal
distribution for proteins with low expression levels and almost similar
fits for proteins with high expression levels. We analysed our GFP
expression data to compare the fits using lognormal and gamma distributions.
For all the three sets of data (\emph{gfp}
fused with the promoters of \emph{mprA},
\emph{sigE} and \emph{rel}),
the lognormal and gamma distribution give similar fits at the different
time points. Figure S6 (Additional File 1) shows a comparison of the
fits for the case of \emph{gfp-mprA}.
The lognormal appears to give a somewhat better fit than the gamma
distribution, specially at the tail ends.

\subsection*{Hysteresis in \textit{gfp} expression}

Some bistable systems exhibit hysteresis, i.e., the response of the
system is history-dependent. In the earlier study, experimental evidence
of hysteresis was obtained with \textit{gfp} fused to the promoter
of \textit{rel}. The experimental procedure followed for the observation
of hysteresis is as follows. In PPK-KO, the \textit{ppk1} knockout
mutant, the \textit{ppk1} gene was introduced under the control of
the\textit{ tet} promoter. We grew PPK-KO carrying the tetracycline-inducible
\textit{ppk1} and\textit{ rel-gfp} plasmid in medium with increasing
concentration of tetracycline (inducer). For each inducer concentration,
the distribution of cells expressing \textit{gfp} was analysed by
flow cytometry in the stationary phase (steady state) and the mean
GFP level was measured. A similar set of experiments was carried out
for decreasing concentrations of tetracycline. In the present study,
hysteresis experiments in the manner described above were carried
out in the two cases of \textit{gfp }fused to \textit{mprA }and \textit{sigE}
promoters respectively.
\begin{figure}
\begin{center}
\includegraphics[scale=1]{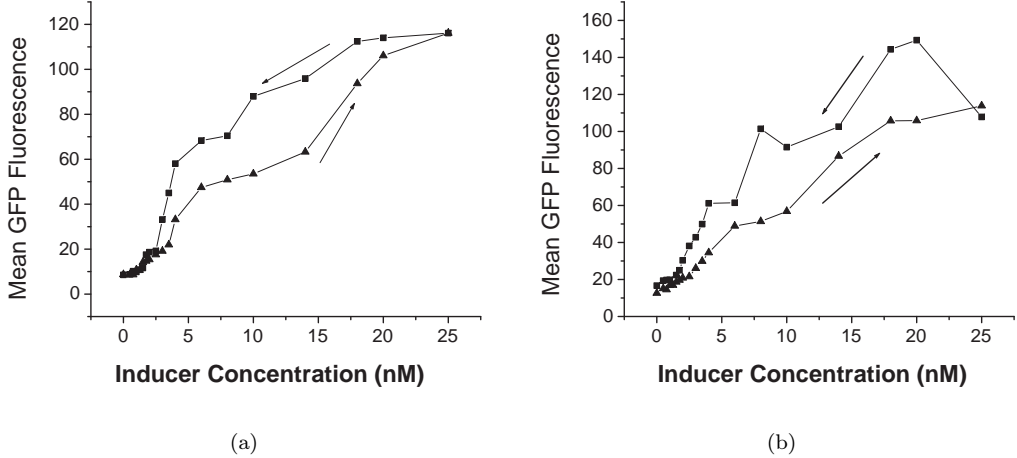}
\end{center}
\caption{\textbf{Hysteresis in }\textbf{\textit{gfp}}\textbf{
expression}\textbf{\textit{.}} The gene\textit{ gfp} is fused with
(a) \textit{mprA} and (b) \textit{sigE} promoter. Filled triangles
and squares represent the experimental data of mean GFP fluorescence
with increasing and decreasing concentrations of tetracycline inducer
respectively.} 
\end{figure}
Figure 7 shows the hysteresis data (mean GFP
fluorescence versus inducer concentration) in the two cases for increasing
(branch going up) and decreasing (branch going down) inducer concentrations.
The existence of two distinct branches is a confirmation of hysteresis
in agreement with theoretical predictions (Figures S1 A-C).
\begin{figure}
\begin{center}
\includegraphics[scale=1]{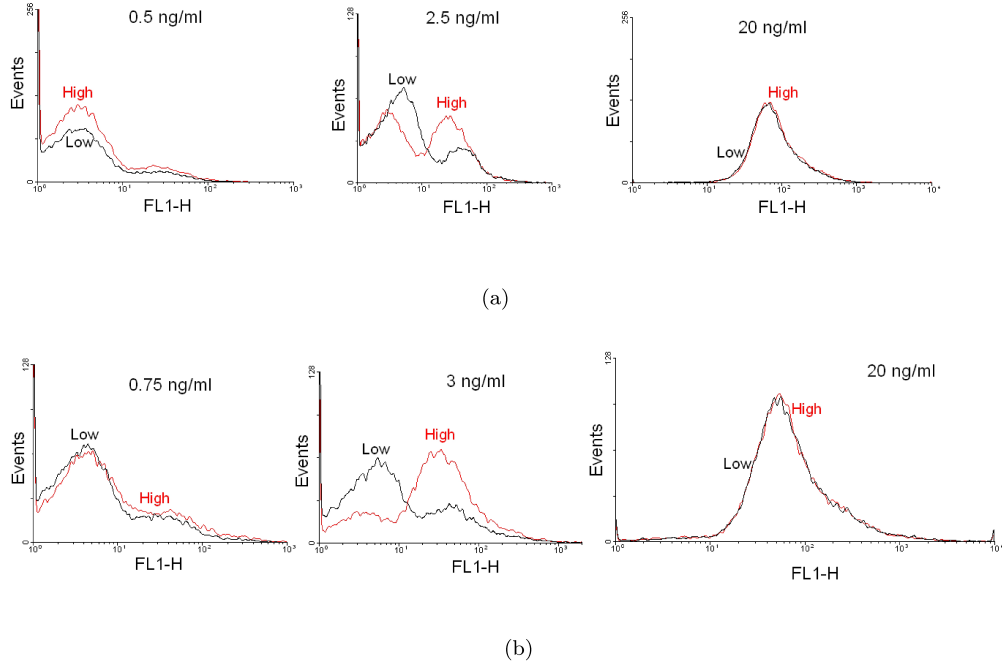}
\end{center}
\caption{\textbf{Hysteresis via GFP distributions. }The
distributions in the stationary phase with two different histories
(see text) when \textit{gfp} is fused with (a) \textit{mprA} and (b)
\textit{sigE} promoter. The specific inducer concentrations are mentioned
with each plot.} 
\end{figure}
 Figure 8 shows the GFP distributions in the stationary phase for two sets
of experiments with different histories, one in which the inducer
concentration is increased from low to a specific value (indicated
as {}``Low'' in black) and the other in which the same inducer concentration
is reached by decreasing the inducer concentration from a high value
(indicated as {}``High'' in red). The distributions show that two
regions of monostability are separated by a region of bistability.
In the cases of monostability, the distributions with different histories
more or less coincide. In the region of bistability, the distributions
are distinct indicating a persistent memory of initial conditions.

\section*{Discussion}

The development of persistence in microbial populations subjected
to stress has been investigated extensively in microorganisms like
\textit{E. coli} and mycobacteria \cite{9,11,21,22,28,29}. In an earlier
study \cite{19}, we demonstrated the roles of positive feedback and
gene expression noise in generating phenotypic heterogeneity in a
population of \textit{M. smegmatis} subjected to nutrient depletion.
The heterogeneity was in terms of two distinct subpopulations designated
as L and H subpopulations. The subpopulations corresponded to persister
and non-persister cell populations with the stringent response being
initiated in the former. In the present study, we have undertaken
a comprehensive single cell analysis of the expression activity of
the three key molecular players in the stringent response pathway,
namely, MprA, SigE and Rel. This has been done by fusing \emph{gfp}
to the respective genes in separate experiments and monitoring the
GFP levels in a population of cells via flow cytometry. The distribution
has been found to be bimodal in each case. 

In our earlier study \cite{19}, with only the positive autoregulation
of the \textit{mprAB} operon taken into account, bistability was obtained
in a parameter regime with restricted experimental relevance. The
inclusion of the effective positive feedback loop due to growth retardation
by protein synthesis gives rise to a considerably more extended region
of bistability in parameter space. The persister cells with high stringent
response regulator levels are known to have slow growth rates \cite{21,22,28,29}.
This is consistent with the view that stress response diverts resources
from growth to stress-related functions resulting in the slow growth
of stress-resistant cells \cite{36}. Figures 2 and 3 provide experimental
evidence that the mean intensity of GFP fluorescence monitoring \emph{mprA-gfp}
expression increases with time while the specific growth rate $\mu$
of the \textit{M. smegmatis} population decreases in the same time interval.
The reciprocal relationship between the two quantities is represented
by an expression similar to that in Eq. (1). Since our knowledge of
the detailed genetic circuitry involved in the stringent response
is limited, we have not attempted to develop a model to explain the
origin of cell growth retardation due to protein synthesis. Further
experiments (e.g., sorting of the mycobacterial cell population into
two subpopulations) are needed to provide conclusive evidence that
increased protein synthesis retards cell growth. The stringent response
pathway involving MprA and MprB is initiated when the mycobacterial
population is subjected to stresses like nutrient depletion. There
is now experimental evidence of complex transcriptional, translational,
and posttranslational regulation of SigE in mycobacteria \cite{46,47,48,49}.
 A double positive feedback loop arises due to the activation
of transcription initiation of \emph{sigE} by MprA-P and the activation
of the transcription of the \emph{mprAB} operon by the SigE-RNAP complex.
Posttranslational regulation of SigE is mediated by RseA, an anti-sigma
factor. Barik et al. \cite{49} have identified a novel positive feedback
involving SigE and RseA which becomes functional under surface stress.
More experiments need to be carried out to obtain insight on the intricate
control mechanisms at work when mycobacteria are subjected to stresses
like nutrient deprivation. This will lead to a better understanding
of the major contributory factors towards the generation of phenotypic
heterogeneity in mycobacterial populations subjected to stress.

%%%%%%%%%%%%%%%%%%%%%%
\section*{Conclusions}
  In the present study, we have characterised quantatively the single
cell promoter activity of three key genes in the stringent response
pathway of the mycobacterial population \emph{M. smegmatis}. Under
nutrient depletion, a {}``responsive switching'' occurs from the
L to the H subpopulation with low and high expression levels respectively.
A comprehensive analysis of the flow cytometry data demonstrates the
coexistence of two subpopulations with overlapping protein distributions.
We have further established that the GFP distribution at any time
point is a linear superposition of a Gaussian and a lognormal distribution.
The coefficients in the linear combination depend on time whereas
the component distributions are time-invariant. The Gaussian and lognormal
distributions describe the distribution of protein levels in the L
and H subpopulations respectively. The two distributions overlap in
a range of GFP fluorescence intensity values. We
also find that the experimental data for the H subpopulation can be
fitted very well by the gamma distribution though the lognormal distribution
gives a slightly better fit. In the case of skewed positive data sets,
the two distributions are often interchangeable \cite{50}. An analytical
framework similar to that in Ref. \cite{43} is, however, yet to be
developed for the mycobacterial stringent response pathway studied
in the paper. The major components in the pathway are the two-component
system $mprAB$ and multiple
positive feedback loops. The two-component system is known to promote
robust input-output relations \cite{51} and persistence of gene expression
states \cite{52} which may partly explain the good fitting of the experimental
data by well-known distributions. Further quantitative measurements
combined with appropriate stochastic modeling are needed to characterise
the experimentally observed subpopulations more uniquely. We used
the binning algorithm developed in \cite{42} to partition the experimental
cell population into the L and H subpopulations. This enabled us to
compute quantities like the mean protein level in the L subpopulation,
the fraction of cells in the H subpopulation and the CV of GFP levels
in the L subpopulation as a function of time. The picture that emerges
from the analysis of experimental data is that of bistability, i.e.,
the coexistence of two distinct subpopulations and stochastic transitions
between the subpopulations resulting in the time evolution of the
fraction of cells in the H subpopulation. As pointed out in the earlier
study \cite{19}, the rate of transition to the H subpopulation and
the CV of the L subpopulation levels attain their maximum values around
the same time point (Figure 5(c)) indicating the role of gene expression
noise in bringing about the transition from the L to the H subpopulation.
We have not attempted to develop theoretical models describing the
time evolution of the relative weights, $\omega_{i}$'s ($i=1,2$),
of the two subpopulations (Eq. (9)). A simple model of two interacting
and evolving subpopulations with linear first order kinetics \cite{12},
cannot explain the sigmoidal nature of the time evolution. A model
with nonlinear growth kinetics has been proposed in \cite{42} but lacking
definitive knowledge on the origin of nonlinearity in the growth of
mycobacterial subpopulations we defer the task of model building to
a future publication.

%%%%%%%%%%%%%%%%%%
\section*{Methods}

\subsection*{Strains }

\emph{M. smegmatis} mc$^{2}$155 was grown routinely in Middle Brook
(MB) 7H9 broth (BD Biosciences) medium supplemented with 2\% glucose
and 0.05\% Tween 80.

\subsection*{Construction of plasmids for fluorescence measurements }

The \emph{mprAB} promoter was amplified from the genomic DNA of \emph{M.
smegmatis} using the sense and antisense primers, 5\textquoteright{}-AA\textbf{GGTACC}GCGCAACACCACAAAAAGCG-3\textquoteright{}
and 5\textquoteright{}-TA\textbf{GGATCC}AGTTTTGACTCACTATCTGAG-3\textquoteright{}
respectively and cloned into the promoter-less replicative\emph{ gfp
}vector pFPV27 between the KpnI and BamHI sites (in bold). The \emph{sigE}
and \emph{rel} promoters fused to \emph{gfp} have been described earlier
\cite{19,31}. The resulting plasmids were electroporated into \emph{M.
smegmatis} mc$^{2}$155 for further study. For the study of hysteresis,
expression of \emph{ppk1} under a tetracycline-inducible promoter
in an \emph{M. smegmatis} strain inactivated in the \emph{ppk1} gene
(PPK-KO), has been described earlier \cite{19}.

\subsection*{FACS analysis}

\emph{M. smegmatis} cells expressing
different promoters fused to GFP were grown in medium supplemented
with kanamycin (25 $\mu g/ml$)
and analysed at different points of time on a FACS Caliber (BD Biosciences)
flow cytometer as described earlier \cite{19}. Briefly, cells were
washed, resuspended in PBS and fluorescence intensity of 20,000 events
was measured. The data was analyzed using Cell Quest Pro (BD Biosciences)
and WINMIDI software. The flow cytometry data is represented in histogram
plots where the x-axis is a measure of fluorescence intensity and
the y-axis represents the number of events.

\subsection*{Measurement of growth rate}

\emph{M. smegmatis } expressing
promoter-\emph{gfp} fusion constructs
were grown in Middle Brook (MB) 7H9 broth supplemented with glucose
and Tween 80, and kanamycin (25 \textit{$\mu g/ml$}).
Growth at different time points was measured by recording absorbance
values at 600 nm (a value of 1 OD$_{600}$ is equal to $10^{8}$ cells
or 200 $\mu g$ dry weight of cells). A growth curve was generated
by plotting absorbance values against time (inset of Figure 2). The
specific growth rate $\mu$ (Eq. (2)) at different time points is
determined by taking derivatives of the growth curve at the different
time points (Figure 2).

%%%%%%%%%%%%%%%%%%%%%%%%%%%%%%%%
\section*{Authors contributions}
 IB, JB and MK conceptualised, supervised and coordinated the study.
KS, JB and MK carried out the experiments. SG, BG and IB developed
the theoretical model, performed the data analysis and interpreted
the data. IB drafted the manuscript. All authors read and approved
of the final version.

%%%%%%%%%%%%%%%%%%%%%%%%%%%
\section*{Acknowledgements}

  IB, JB and MK thank M. Thattai for some useful discussions. This work
was supported in part by a grant from the Department of Biotechnology,
Government of India to MK. SG is supported by CSIR, India, under Grant
No. 09/015(0361)/2009-EMR-I.

\pagebreak{}

\pagebreak{}
%%%%%%%%%%%%%%%%%%%%%%%%%%%%%%%%%%%%%%%%%%%%%%%%%%%%%%%%%%%%%%%%%%%%%%%%%%%%%%

\part*{Supplementary Information}

\section*{Mathematial model}

The reaction scheme describing the processes shown in Figure 1 is
given by

\begin{equation}
B_{2}+B_{2}\:\begin{array}{c}
k_{1}\\
\rightleftharpoons\:\\
k_{2}\end{array}B_{2}-B_{2}\:\overset{k_{3}}{\longrightarrow\:}B_{1}+B_{1}
\end{equation}

\begin{equation}
A_{2}+B_{1}\begin{array}{c}
k_{4}\\
\:\rightleftharpoons\:\\
k_{5}\end{array}A_{2}-B_{1}\overset{k_{6}}{\:\longrightarrow}\: A_{1}+B_{2}
\end{equation}

\begin{equation}
A_{1}+B_{2}\begin{array}{c}
k_{7}\\
\:\rightleftharpoons\\
k_{8}\end{array}\: A_{1}-B_{2}\overset{k_{9}}{\:\longrightarrow\:}A_{2}+B_{2}
\end{equation}

\begin{equation}
G_{AB}+A_{1}\begin{array}{c}
k_{a}\\
\:\rightleftharpoons\:\\
k_{d}\end{array}G_{AB}^{*}\overset{\beta}{\:\longrightarrow\:}A_{2}+B_{2}
\end{equation}

\begin{equation}
G_{AB}+A_{1}\overset{s}{\:\longrightarrow\:}A_{2}+B_{2}
\end{equation}

In the equations, $A_{1}(A_{2})$ represents the phosphorylated (unphosphorylated)
form of MprA and $B_{1}(B_{2})$ denotes the phosphorylated (unphosphorylated)
form of MprB. The inactive and active states of the \textit{mprAB}
operon are represented by $G_{AB}$ and $G_{AB}${*} respectively.
In the inactive state, MprA and MprB proteins are synthesized at a
basal rate s and in the active state protein production occurs at
an enhanced rate $\beta$. Eq. (1) describes
the autophosphorylation reaction of MprB with the poly P chain serving
as a source of phosphate groups. Eq. (2) describes the transfer of
the phosphate group from the phosphorylated MprB to MprA. Eq. (3)
corresponds to dephosphorylation of phosphorylated MprA by unphosphorylated
MprB which thus acts as a phosphatase (in the earlier study \cite{key-1},
phosphorylated MprB was assumed to act as a phosphatase which is not
consistent with experimental evidence). Eqs. (4) and (5) describe
activation of the \textit{mprAB} operon by phosphorylated MprA and
basal expression of the operon respectively. Refs. \cite{key-2,key-3,key-4} provide
experimental justification for the reaction scheme shown in Eqs. (1)-(5).
Using standard mass action kinetics, we write down the rate equations
for the concentration of each of the key molecular species participating
in the biochemical events. The equations are: 

\begin{equation}
\frac{d[A_{1}]}{dt}\:=
\: k_{6}[A_{2}-B_{1}]-k_{7}[A_{1}][B_{2}]+k_{8}[A_{1}-B_{2}]-\gamma[A_{1}]-
\frac{\phi[A_{1}]}{1+\theta_{1}[A_{1}]}
\end{equation}

\begin{equation}
\frac{d[A_{2}]}{dt}\:=
\: s+\beta\frac{[A_{1}]/k}{1+[A_{1}]/k}+k_{9}[A_{1}-B_{2}]-k_{4}[A_{2}][B_{1}]+k_{5}[A_{2}-B_{1}]
-\gamma_{1}[A_{2}]-\frac{\phi[A_{2}]}{1+\theta_{2}[A_{2}]}
\end{equation}

\begin{equation}
\frac{d[B_{1}]}{dt}\:=\: k_{3}[B_{2}-B_{2}]-k_{4}[A_{2}][B_{1}]+k_{5}[A_{2}-B_{1}]-\gamma[B_{1}]-\frac{\phi[B_{1}]}{1+\theta_{1}[B_{1}]}
\end{equation}

\begin{equation}
\frac{d[B_{2}]}{dt}\:=\: s+\beta\frac{[A_{1}]/k}{1+[A_{1}]/k}+k_{2}[B_{2}-B_{2}]-k_{1}[B_{2}]^{2}+k_{6}[A_{2}-B_{1}]-k_{7}[A_{1}][B_{2}]+(k_{8}+k_{9})[A_{1}-B_{2}]-\gamma[B_{2}]-\frac{\phi[B_{2}]}{1+\theta_{2}[B_{2}]}\end{equation}

\begin{equation}
\frac{d[B_{2}-B_{2}]}{dt}\:=\:-k_{2}[B_{2}-B_{2}]+k_{1}[B_{2}]^{2}-k_{3}[B_{2}-B_{2}]\end{equation}

\begin{equation}
\frac{d[A_{2}-B_{1}]}{dt}\:=\: k_{4}[A_{2}][B_{1}]-k_{5}[A_{2}-B_{1}]-k_{6}[A_{2}-B_{1}]\end{equation}

\begin{equation}
\frac{d[A_{1}-B_{2}]}{dt}\:=\: k_{7}[A_{1}][B_{2}]-(k_{8}+k_{9})[A_{1}-B_{2}]\end{equation}

\begin{equation}
\frac{d[SigE]}{dt}\:=\: s_{1}+\beta_{1}\frac{[A_{1}]/k'}{1+[A_{1}]/k'}-\delta_{1}[SigE]\end{equation}

\begin{equation}
\frac{d[GFP]}{dt}\:=\: s_{2}+\beta_{2}\frac{[SigE]/k''}{1+[SigE]/k''}-\delta_{2}[GFP]\end{equation}

Eq. (13) represents SigE synthesis due to transcriptional activation
of the\textit{ sigE }gene by phosphorylated MprA-P. Eq. (14) describes
GFP production due to the activation of the \textit{rel} promoter
by SigE. The rate constants $\gamma$, $\gamma_{1}$, $\delta_{1}$
and $\delta_{2}$ are the degradation rate constants.
The last terms in Eqs. (6)-(9) represent the nonlinear decay rates
the genesis of which is explained in the main text (see Eq. 4) \cite{key-5}.
Eqs. (6)-(14) correspond to the case where\textit{ gfp} is fused to
the\textit{ rel} promoter. In the other cases when \textit{gfp} is
fused to the \textit{mprA} or \textit{sigE }promoter, appropriate
modifications in the set of equations are required.

The steady state solution of Eqs. (6)-(14) is obtained by setting
all the rates of change to be zero. In the case of bistability, there
are three steady state solutions, two stable and one unstable \cite{key-6,key-7,key-8}.
In the steady state, one has to solve the following set of coupled
nonlinear algebraic equations:

\begin{equation}
\alpha_{1}[A_{2}][B_{1}]-\alpha_{2}[A_{1}][B_{2}]-\gamma[A_{1}]-\frac{\phi[A_{1}]}{1+\theta_{1}[A_{1}]}=0\end{equation}

\begin{equation}
s+\beta\frac{[A_{1}]/k}{1+[A_{1}]/k}-\alpha_{1}[A_{2}][B_{1}]+\alpha_{2}[A_{1}][B_{2}]-\gamma_{1}[A_{2}]-\frac{\phi[A_{2}]}{1+\theta_{2}[A_{2}]}=0\end{equation}

\begin{equation}
\alpha[B_{2}]^{2}-\alpha_{1}[A_{2}][B_{1}]-\gamma[B_{1}]-\frac{\phi[B_{1}]}{1+\theta_{1}[B_{1}]}=0\end{equation}

\begin{equation}
s+\beta\frac{[A_{1}]/k}{1+[A_{1}]/k}-\alpha[B_{2}]^{2}+\alpha_{1}[A_{2}][B_{1}]-\gamma[B_{2}]-\frac{\phi[B_{2}]}{1+\theta_{2}[B_{2}]}=0\end{equation}

\begin{equation}
s_{1}+\beta_{1}\frac{[A_{1}]/k'}{1+[A_{1}]/k'}-\delta_{1}[SigE]=0\end{equation}

\begin{equation}
s_{2}+\beta_{2}\frac{[SigE]/k''}{1+[SigE]/k''}-\delta_{2}[GFP]=0\end{equation}

where,\begin{equation}
\alpha=\frac{k_{1}k_{3}}{k_{2}+k_{3}},\;\alpha_{1}=\frac{k_{4}k_{6}}{k_{5}+k_{6}},\;\alpha_{2}=\frac{k_{7}k_{9}}{k_{8}+k_{9}},\; k=\frac{k_{d}}{k_{a}}\end{equation}

The solutions of Eqs. (15)-(20) are obtained with the help of Mathematica.
Figures S1 A-C show the steady state solutions generated by varying
the parameter $\alpha$ (associated with the autophosphorylation of
MprB). The parameters have values:$\alpha_{1}=2.4,\alpha_{2}=2.8,\gamma=0.1,s=0.14,\beta=4,k=1,\gamma_{1}=1,\phi=0.5,\theta_{1}=1,\theta_{2}=10,s_{1}=0.02,\beta_{1}=4,k'=10,\delta_{1}=1,$
$s_{2}=0.12,\beta_{2}=4,k''=2,\delta_{2}=0.1$ in appropriate units.

In each of the Figures S1 A-C, the solid and dotted branches represent
stable and unstable steady states respectively. Bistability is obtained
over a wide range of parameter values due to the inclusion of the
non-linear decay term in Eqs. (6)-(9). In the hysteresis experiments,
the inducer tetracycline was used to control the level of PPK1 and
therefore the synthesis of the poly P chain. Since the latter acts
as the source of phosphate groups for the autophosphorylation of MprB
(Eq. (1)), the rate constant k, in Eq. (1) is effectively proportional
to the inducer (or the PPK1) concentration. As the parameter $\alpha$
(Eq. (21)) includes the rate constant $k_{1}$ , a varying inducer
concentration is equivalent to varying the parameter $\alpha$. There
is some experimental evidence that MprA-P regulates the expression
of the \textit{mprAB} operon in the form of dimers \cite{key-9}. Inclusion
of this feature in our model makes the bistable behaviour more prominent.

\begin{figure}[H]
\begin{center}
\includegraphics[scale=1]{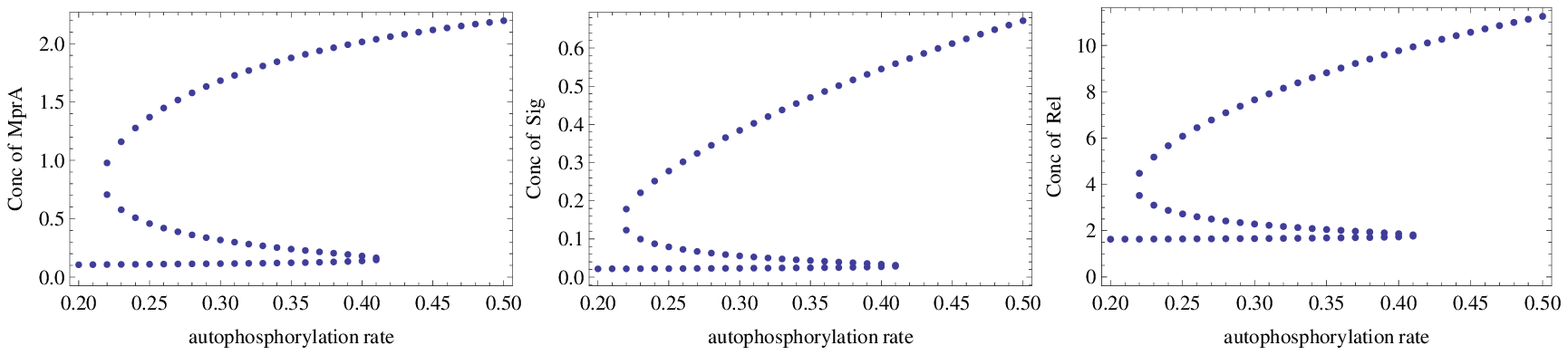}
\end{center}
\textbf{Figure S1:} Bistability and hysteresis in the deterministic model.
Steady state concentrations of MprA, SigE and Rel versus the parameter
$\alpha$ (Eq. (21)).
\end{figure}

\begin{figure}[H]
\begin{center}
\includegraphics[scale=1]{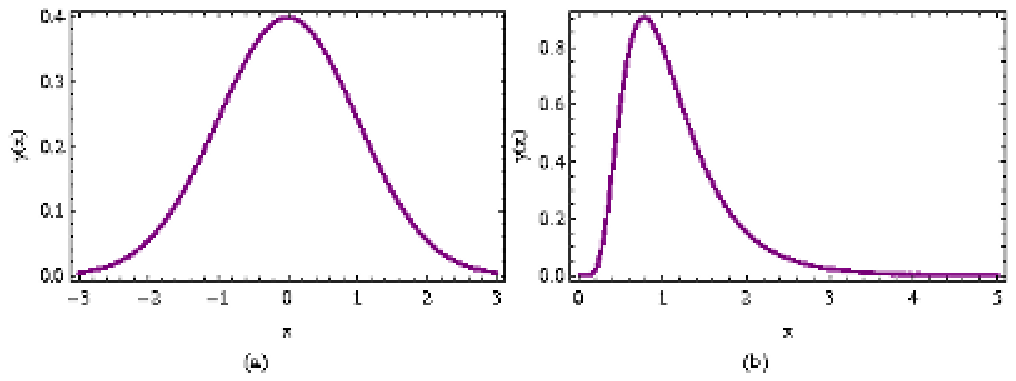}
\end{center}
\textbf{Figure S2:} (a) Gaussian and (b) lognormal distributions which describe
the distribution of GFP leads in the L and H subpopulations respectively.
\end{figure}

\begin{figure}[H]
\begin{center}
\includegraphics[scale=1]{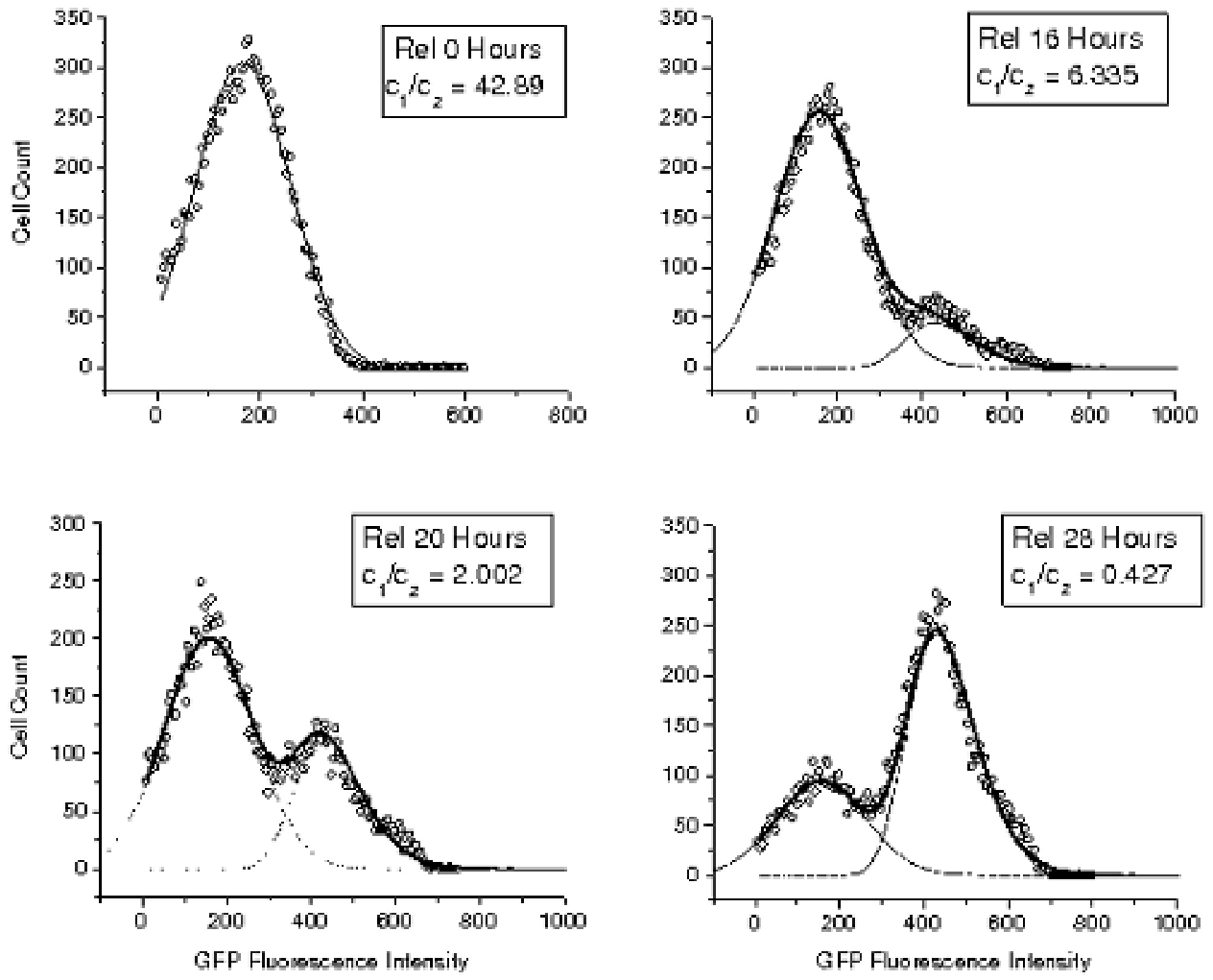}
\end{center}
\textbf{Figure S3:} Experimental data for cell count versus GFP fluorescence
intensity at selected time points when \textit{gfp} is fused with
\textit{rel} promoter. The solid curve represents $P(x,t)$ in Eq.
(5) and the dotted curves are the individual terms on the r.h.s. The
different parameters of $P_{1}(x)$ and $P_{2}(x)$ have the values
$x_{01}=157.14748,\; w_{01}=150.43575,\; x_{02}=6.10036,\; w_{02}=0.1847$
when \textit{gfp} is fused with \textit{rel}.
\end{figure}

\pagebreak

\begin{figure}[H]
\begin{center}
\includegraphics[scale=1]{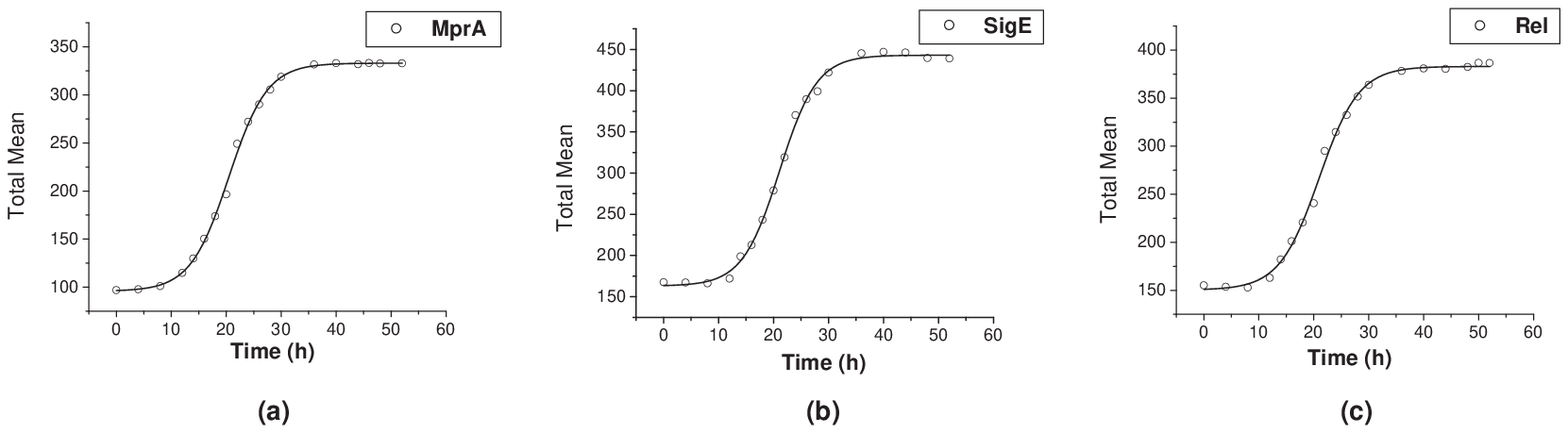}
\end{center}
\textbf{Figure S4:} Mean GFP fluorescence level for the total population versus
time in the three cases of \emph{gfp} fused with the promoters of
(a) \textit{mprA}, (b) \textit{sigE} and (c) \textit{rel}. 
\end{figure}

\begin{figure}[H]
\begin{center}
\includegraphics[scale=1]{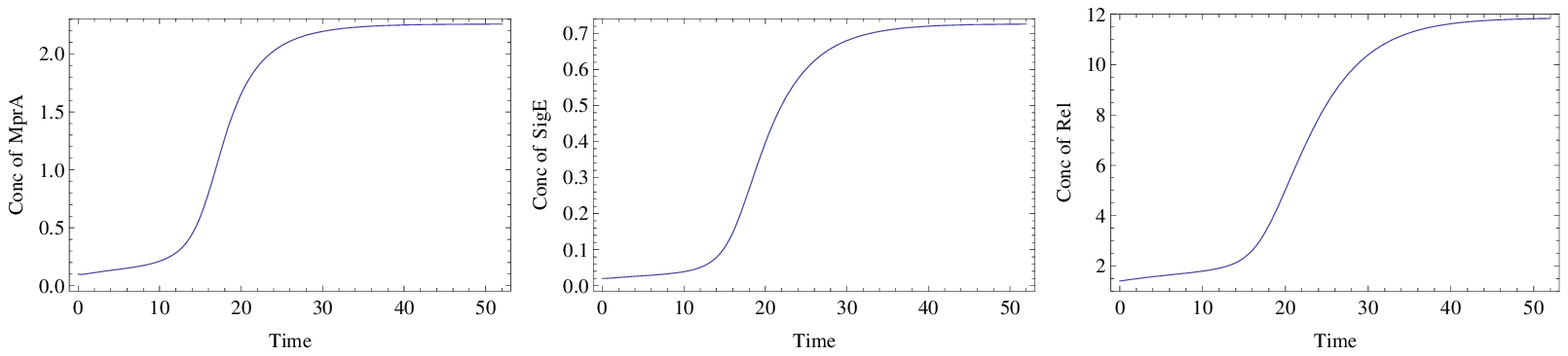}
\end{center}
\textbf{Figure S5:} Concentrations of MprA, SigE and GFP (in arbitrary units)
versus time. The values of the concentrations are obtained by solving
Eqs. (6)-(14) in Text S1.
\end{figure}

\pagebreak

\begin{figure}[H]
\begin{center}
\includegraphics[scale=0.9]{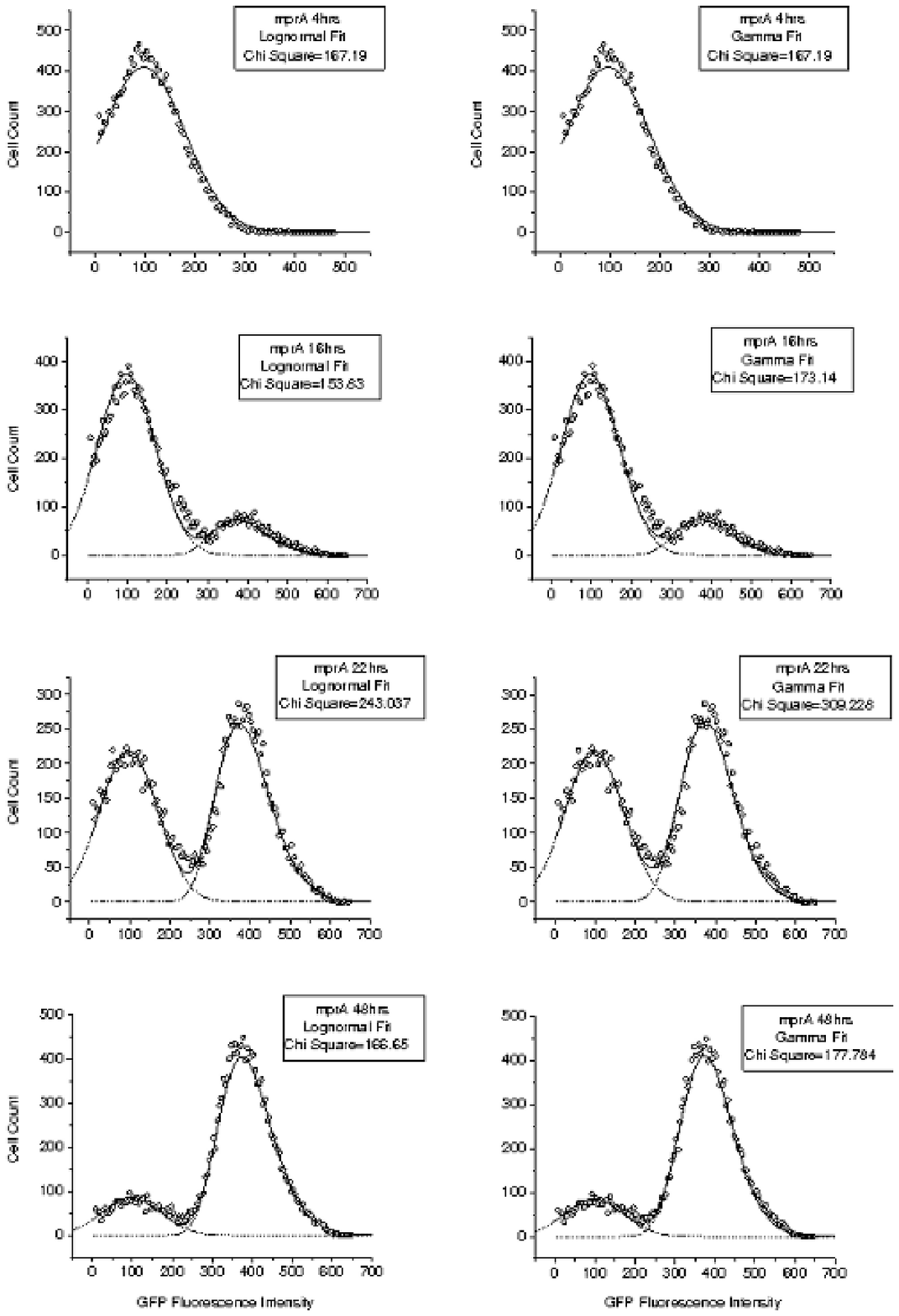}
\end{center}
\textbf{Figure S6:} Comparison of fits of experimental data
for cell count versus GFP fluorescence intensity at selected time
points when \emph{gfp} is fused
with \emph{mprA} promoter, with
lognormal (Eq. (7)) and gamma distributions. The gamma distribution
has the form $P(x)=\frac{x^{a-1}\, exp(-\frac{x}{b})}{b^{a}\,\Gamma(a)}$
, where the parameters \emph{a}
and \emph{b} have the values $a=33.246,$
$b=11.59$ and $\Gamma(a)$ is the gamma function.

\end{figure}

\end{document}